\documentclass{Rinton-P9x6}

\begin{document}

\title{Monopole Problem and Extensions of Supersymmetric
 Hybrid Inflation}
\author{R. Jeannerot}
\address{SISSA, Via Beirut 2--4, 34013 Trieste, Italy and INFN,
sez. di Trieste, Trieste, Italy}

\author{S. Khalil}
\address{
Centre for Theoretical Physics, University of Sussex, Brighton BN1
9QJ,~~U.~K.\\
Ain Shams University, Faculty of Science, Cairo, 11566, Egypt.}
\author{G. LAZARIDES}

\address{Physics Division, School of Technology, 
Aristotle University of Thessaloniki, 
Thessaloniki GR 540 06, Greece}

%%%%%%%%%%%%%%%%%%%%%%%%%%%%%%%%%%%%%%%%%%%%%%%%%%%%%%%%%%%%%%
% You may repeat \author \address as often as necessary %
%%%%%%%%%%%%%%%%%%%%%%%%%%%%%%%%%%%%%%%%%%%%%%%%%%%%%%%%%%%%%%

\maketitle

\abstracts{
We discuss, in the context of a concrete supersymmetric grand 
unified model based on the Pati-Salam gauge group $SU(4)_c\times 
SU(2)_L\times SU(2)_R$~, two `natural' extensions of supersymmetric 
hybrid inflation, which avoid the cosmological disaster 
encountered in the standard hybrid inflationary scenario from the 
overproduction of monopoles at the end of inflation. Successful 
`reheating' which satisfies the gravitino constraint takes place 
after the end of inflation. Also, adequate baryogenesis via a
primordial leptogenesis occurs consistently with the solar and
atmospheric neutrino oscillation data as well as the $SU(4)_c$
symmetry. Moreover, the $\mu$-term is generated via a 
Peccei-Quinn symmetry and proton is practically stable.}

\section{Introduction} 
\label{sec:introduction}

Inflation offers an elegant solution to the outstanding problems of
the standard Big-Bang cosmological model and predicts the formation 
of the large scale structure of the universe and the temperature 
fluctuations which are observed in the cosmic microwave background 
radiation (CMBR). It also solves the cosmological problem caused by 
the overproduction of grand unified theory (GUT) magnetic monopoles 
as well as other unwanted relics such as domain walls, gravitini or 
moduli fields.

However, the early realizations of inflation require extremely flat
potentials and very small coupling constants. To solve this 
naturalness problem, the hybrid inflationary scenario has been 
introduced \cite{linde}. The basic idea was to use two real scalar 
fields $\chi$ and $\sigma$ instead of one that was normally used. 
The field $\chi$ may be a gauge non-singlet and provides the `vacuum' 
energy density which drives inflation, while $\sigma$ is the slowly 
varying field during inflation. This splitting of roles between two 
fields allows us to reproduce the observed temperature fluctuations 
of the CMBR with `natural' (not too small) values of the relevant 
parameters in contrast to previous realizations of inflation. 

The scalar potential for hybrid inflation possesses a valley of 
local minima with respect to $\chi$ with large `vacuum' energy 
density. This valley lies at $\chi=0$ with $\sigma$ being greater 
than a certain critical (instability) value $\sigma_c$, and has 
a classical inclination provided by the mass of $\sigma$. The 
global minima of the potential lie at $\chi\neq 0$ and $\sigma=0$. 
As the system rolls down the valley of local minima, the slow-roll 
conditions (see e.g., Ref.\cite{cosmology}) are satisfied and 
inflation takes place. Inflation ends abruptly as $\sigma$ falls 
below $\sigma_c$. It is followed by a `waterfall' regime and 
$\chi$ starts oscillating about a global minimum of the potential 
acquiring a non-vanishing vacuum expectation value (vev). If 
$\chi$ is a gauge non-singlet, spontaneous gauge symmetry 
breaking occurs at the end of inflation, and topological defect 
can potentially form \cite{smooth}.

The simplest framework for realizing hybrid inflation is provided
\cite{Cop,dvasha} by supersymmetric (SUSY) GUTs which are based on
gauge groups with rank greater than five. The same superpotential
which lowers the rank of the gauge group also leads \cite{dvasha} 
to successful hybrid inflation with `natural' values of the 
relevant parameter and a gauge symmetry breaking scale of the order 
of the SUSY GUT scale. The slowly rolling inflaton field belongs to 
a gauge singlet superfield which couples to a conjugate pair of 
gauge non-singlet Higgs superfields. The tree-level scalar potential 
possesses a flat valley of local minima for values of the gauge 
singlet inflaton greater than a certain critical value. Along this 
valley, the vevs of the Higgs superfields vanish, there exists 
a constant non-zero `vacuum' energy density and SUSY is broken. 
The (classical) flatness of the valley is lifted by the one-loop 
radiative corrections \cite{dvasha} to the scalar potential 
which are calculated with the GUT gauge symmetry being restored 
and SUSY being broken. A variant of Linde's scenario is thus 
obtained. Inflation ends by a `waterfall' regime as the gauge 
singlet falls below its critical value, the Higgs fields and the 
gauge singlet start oscillating about the SUSY minima of the 
potential where the Higgs vevs are non-zero. 

If the SUSY vacuum manifold is homotopically non-trivial, 
topological defects will be copiously formed \cite{smooth} by 
the Kibble mechanism \cite{kibble} since the system can end up 
at any point of the vacuum manifold with equal probability. So 
a cosmological disaster is encountered in the hybrid inflationary 
models which are based on a gauge symmetry breaking which predicts 
the existence of magnetic monopoles. One way out of this 
catastrophe is to do this symmetry breaking in two steps by 
introducing an intermediate symmetry breaking scale between the GUT 
and the standard model scales. The intermediate gauge symmetry must 
be chosen such that the unwanted monopoles are formed in the 
first step of symmetry breaking, and hybrid inflation occurs in 
the second step which does not lead to the formation of new 
unwanted topological defects. Inflation then dilutes the 
pre-existing monopoles without generating new ones. The rank of 
the gauge group must be lowered in the second step and, in many 
realistic GUTs, cosmic strings are formed at the end of inflation 
\cite{RJSSB}. They will contribute to the CMBR anisotropy in a 
proportion which depends upon the GUT gauge group and the cosmic 
microwave explorer (COBE) \cite{cobe} normalization for strings 
and inflation.

One idea \cite{smooth,jean1,jean2} for solving the monopole 
problem of hybrid inflation is to include into the standard 
superpotential for hybrid inflation the leading 
non-renormalizable term. This term, as we will explain in the 
next section, cannot be excluded by any symmetries and, if its 
dimensionless coefficient is of order unity, can be comparable 
with the trilinear coupling of the standard superpotential 
(whose coefficient is $\sim 10^{-3}$). Actually, we have two 
options. We can either keep \cite{jean1} both these terms or 
remove \cite{smooth,jean2} the trilinear term by imposing an 
appropriate discrete symmetry and keep only the leading 
non-renormalizable term. The pictures which emerge in the two 
cases are quite different. However, they share an important 
common feature. The GUT gauge group is already broken during 
inflation and thus no topological defects can form at the end 
of inflation. Consequently, the monopole problem is solve even 
in GUTs with a single step of symmetry breaking.

Furthermore, the constraints on the quadrupole anisotropy of the 
CMBR from the COBE \cite{cobe} measurements can be easily 
satisfied. Our model possesses a number of other interesting 
features too. The $\mu$ problem of the minimal supersymmetric 
standard model (MSSM) is solved \cite{PQmu} via a Peccei-Quinn 
(PQ) symmetry \cite{pq} which also solves the strong CP problem. 
Although the baryon and lepton numbers are explicitly 
violated, the proton life time is considerably higher than the 
present experimental limits. Light neutrinos acquire hierarchical 
masses by the seesaw mechanism and the baryon asymmetry of the 
universe (BAU) can be generated via a primordial leptogenesis 
\cite{leptogenesis} (for a recent review see Ref.\cite{springer}). 
The gravitino constraint \cite{khlopov} 
on the `reheat' temperature, the low deuterium abundance limits 
\cite{deuterium} on the BAU and the requirement of almost 
maximal $\nu_{\mu}-\nu_{\tau}$ mixing from SuperKamiokande 
\cite{japan} can be met for $\mu$- and $\tau$-neutrino masses 
restricted by the small or large mixing angle MSW solution 
\cite{bahcall} of the solar neutrino puzzle and SuperKamiokande 
respectively. The required values of the relevant parameters are 
`natural'.

\section{SUSY Hybrid Inflation and its Extensions}
\label{sec:hybrid}

We will now summarize the standard SUSY hybrid inflationary 
scenario in the context of a concrete SUSY GUT and discuss its 
extensions which solve the magnetic monopole problem 
encountered in the standard scenario. Along the lines of 
Refs.\cite{jean1,jean2}, we consider the SUSY Pati-Salam (PS) 
model \cite{ps} which is one of the simplest GUT models 
predicting magnetic monopoles. This model is based on the PS 
gauge group $G_{PS}=SU(4)_c\times SU(2)_L\times SU(2)_R$. 
The PS monopoles carry two units of `Dirac' magnetic charge 
\cite{magg}. We will present possible solutions of the 
magnetic monopole problem of hybrid inflation within the SUSY 
PS model \cite{jean1,jean2}. It is worth mentioning, however, 
that these solutions can be readily applied to other semi-simple 
gauge groups too such as the `trinification' group $SU(3)_c
\times SU(3)_L\times SU(3)_R$, which emerges from string 
theory and predicts \cite{trinification} monopoles with triple 
`Dirac' magnetic charge, and possibly to simple gauge groups 
such as $SO(10)$.

In the SUSY PS model, the left-handed quark and lepton 
superfields are accommodated in the following representations:
\begin{eqnarray}
F_i &=& (4,2,1) \equiv \left(\begin{array}{cccc}
                       u_i & u_i & u_i & \nu_i\\
                      d_i & d_i & d_i & e_i
                      \end{array}\right) , \nonumber\\
F^c_i &=& (\bar{4},1,2) \equiv \left(\begin{array}{cccc}
                       u^c_i & u^c_i & u^c_i & \nu^c_i\\
                      d^c_i & d^c_i & d^c_i & e^c_i
                      \end{array}\right) ,
\end{eqnarray}
where the subscript $i=1,2,3$ denotes the family index. The 
$G_{PS}$ gauge symmetry can be spontaneously broken to the 
standard model gauge group by a pair of Higgs superfields
\begin{eqnarray}
H^c &=& (\bar{4},1,2) \equiv \left(\begin{array}{cccc}
                       u^c_H & u^c_H & u^c_H & \nu_H^c\\
                       d^c_H & d^c_H & d^c_H & e^c_H
                      \end{array}\right) , \nonumber\\
\bar{H}^c &=& (4,1,2) \equiv \left(\begin{array}{cccc}
                       \bar{u}^c_H & \bar{u}^c_H & 
                       \bar{u}^c_H &
                       \bar{\nu}_H^c\\
                      \bar{d}^c_H & \bar{d}^c_H & 
                      \bar{d}^c_H & \bar{e}^c_H
                      \end{array}\right)
\end{eqnarray}
acquiring non-vanishing vevs in the right-handed neutrino 
direction, $\langle\nu_H^c\rangle$, $\langle\bar{\nu}_H^c
\rangle\neq 0$. The two low energy Higgs doublets of the MSSM 
are contained in the following representation:
\begin{eqnarray}
h = (1,2,2) \equiv \left(\begin{array}{cc}
                      h_2^+ & h_1^0 \\
                      h_2^0 & h_1^-
                      \end{array}\right).
\end{eqnarray}
After the breaking of $G_{PS}$, the bidoublet Higgs field $h$ 
splits into two Higgs doublets $h_1$, $h_2$, whose neutral 
components subsequently develop weak vevs $\langle h^0_1\rangle
=v_1$ and $\langle h^0_2\rangle=v_2$ with $\tan\beta=v_2/v_1$. 

The (renormalizable) superpotential for the breaking of $G_{PS}$ 
is
\begin{equation}
W=\kappa S(-M^2+H^c\bar H^{c})~, 
\label{eq:W}
\end{equation}
where $S$ is a gauge singlet left-handed superfield and the 
parameters $\kappa$, $M$ can be made positive by field 
redefinitions. The vanishing of the F-term $F_S$ implies that 
$\langle H^c \rangle\langle\bar{H}^c\rangle=M^2$, whereas 
the D-terms vanish for $\vert\langle H^c\rangle\vert=
\vert\langle\bar{H}^c\rangle\vert$. So, the SUSY vacua 
(rotated to the real axis) lie at 
$\langle H^c\rangle=\langle\bar{H}^c\rangle^*=
\pm M$ and $\langle S\rangle=0$ (from $F_{H^c}=
F_{\bar{H}^c}=0$). We see that $W$ leads to the spontaneous 
breaking of $G_{PS}$.

It is interesting to note that the same superpotential which 
breaks $G_{PS}$ also leads to hybrid inflation. The potential 
derived from $W$ in Eq.(\ref{eq:W}) is
\begin{equation}
V(H^c,\bar{H}^c,S)=
\kappa^2\vert M^2-H^c\bar{H}^c\vert^2+
\kappa^2\vert S\vert^2(\vert H^c\vert^2+
\vert\bar{H}^c\vert^2)+{\rm{D-terms}}.
\label{eq:hybpot}
\end{equation}
For $\vert S\vert>S_{c}\equiv M$, the potential $V$ is 
minimized by $H^c=\bar{H}^c=0$. This yields a classically 
flat valley of local minima. However, the flatness of this 
valley is lifted at the one-loop level. The SUSY breaking by 
the `vacuum' energy density $\kappa^2M^4$ along this valley 
causes a mass splitting in the supermultiplets $H^c$, 
$\bar{H}^c$. We obtain a Dirac fermion with ${\rm mass}^2$ 
equal to $\kappa^2\vert S\vert^2$ and two complex scalars 
with ${\rm mass}^2$ equal to 
$\kappa^2\vert S\vert^2\pm\kappa^2M^2$. This leads to 
the existence of important one-loop radiative corrections to 
$V$ on the valley which can be found from the Coleman-Weinberg 
formula \cite{cw}:
\begin{equation}
\Delta V=\frac{1}{64\pi^2}\sum_i(-)^{F_i}\ M_i^4\ln
\frac{M_i^2}{\Lambda^2}~, 
\label{eq:deltav}
\end{equation}
where the sum extends over all helicity states $i$, $F_i$ and
$M_i^2$ are the fermion number and ${\rm mass}^2$ of the 
$i$th state, and $\Lambda$ is a renormalization mass scale. 
We find that $\Delta V(\vert S\vert)$ is given \cite{lisbon} 
by
\begin{equation}
\kappa^2 M^4~{\kappa^2\over 4\pi^2}\left(
2\ln{\kappa^2\vert S\vert^2\over\Lambda^2}
+(z+1)^{2}\ln(1+z^{-1})+(z-1)^{2}\ln(1-z^{-1})\right),
\label{eq:rc}
\end{equation}
where $z=\vert S\vert^2/M^2$. For $z\gg 1$ 
($\vert S\vert\gg S_c$), the effective potential on the 
valley can be expanded \cite{dvasha,lss} as 
\begin{equation}
V_{{\rm{eff}}}(\vert S\vert)=\kappa^2 M^4
\left[1+\frac{\kappa^2}{2\pi^2}\left(\ln 
\frac{\kappa^2\vert S\vert^2}{\Lambda^2}
+\frac{3}{2}-\frac{1}{12z^2}+\cdots\right)\right].
\label{eq:veff}
\end{equation}

We see that the one-loop radiative corrections generate a 
($\Lambda$-independent) slope along the classically flat 
valley of local minima. So this valley can, in principle, be 
used as an inflationary trajectory. As the system rolls down 
the valley driven by the contribution in Eq.(\ref{eq:veff}), 
the energy density is dominated by the tree-level `vacuum' 
energy density $\kappa^2 M^4$, the slow-roll conditions hold, 
and inflation takes place till $\vert S\vert$ reaches its 
critical value $S_c$. The COBE \cite{cobe} measurements on 
the quadrupole anisotropy of the CMBR can be reproduced 
\cite{dvasha} with `natural' values of $\kappa$, and $M$'s 
close to the SUSY GUT scale. 

At $S_c$, the system enters into a `waterfall' regime followed 
by damped oscillations about the SUSY vacua where $H^c$ and 
$\bar{H}^c$ acquire non-zero vevs and $G_{PS}$ breaks. It 
is an important feature of the scenario that the $G_{PS}$ 
gauge symmetry is restored along the inflationary trajectory 
and breaks spontaneously only at the end of inflation when the 
system falls towards the SUSY minima. This transition then 
leads \cite{smooth} to a cosmologically unacceptable copious 
production of doubly charged magnetic monopoles. One way to 
resolve this problem, which arises if standard hybrid inflation 
is employed, is to use as inflationary trajectory another flat 
direction in which $G_{PS}$ is already broken. Such a direction 
naturally appears if we include the next order non-renormalizable 
superpotential coupling of $S$ to $H^c$, $\bar{H}^c$. The
trilinear term in Eq.(\ref{eq:W}) can be either kept 
\cite{jean1} or removed \cite{smooth,jean2} by a discrete 
symmetry.

\subsection{Shifted Hybrid Inflation}
\label{subsec:shifted}

As mentioned above, the cosmological monopole problem can be 
solved by including the leading non-renormalizable term in the 
superpotential for hybrid inflation. We will first examine the 
case where the trilinear term in Eq.(\ref{eq:W}) is also kept. 
The coexistence of these terms leads \cite{jean1} to the 
appearance of a new `shifted' classically flat direction where 
the $G_{PS}$ gauge symmetry is broken, i.e., the Higgs fields 
$H^c$, $\bar{H}^c$ possess (constant) non-vanishing vevs. The 
trivial valley of minima where $G_{PS}$ is restored is also 
present. The `shifted' flat direction can be used as an 
alternative inflationary trajectory with the necessary 
inclination obtained again from one-loop radiative corrections, 
which now have to be calculated with both the GUT gauge symmetry 
and SUSY being broken. The termination of inflation is again 
abrupt followed by a `waterfall', but no monopoles are formed 
in this transition since $G_{PS}$ is already spontaneously
broken during inflation.

The relevant part of the superpotential, which includes the 
leading non-renormalizable term, is
\begin{equation}
W=\kappa S(-M^2+H^c\bar{H}^c)-
\beta\frac{S(H^c\bar{H}^c)^2}{M_S^2}~, 
\label{eq:susyinfl}
\end{equation}
where $M_S\approx 5\times 10^{17}~{\rm GeV}$ is the string 
scale and $\beta$ is taken positive for simplicity. 
D-flatness implies that $H^c \,^{*}=e^{i\theta}\bar{H}^{c}$. 
We restrict ourselves to the direction with $\theta=0$ 
($H^c \,^{*}=\bar{H}^{c}$) containing the non-trivial 
(`shifted') inflationary path (see below). The scalar potential 
derived from $W$ in Eq.(\ref{eq:susyinfl}) then takes the form
\begin{equation}
V=\left[\kappa(\vert H^c\vert^2-M^2)-\beta\frac{\vert H^c
\vert^4}{M_S^2}\right]^2+2\kappa^2\vert S\vert^2
\vert H^c\vert^2 
\left[1-\frac{2\beta}{\kappa M_S^2}\vert H^c\vert^2
\right]^2.
\label{eq:inflpot}
\end{equation}
Defining the dimensionless variables $y=\vert H^c\vert/M$, 
$w=\vert S\vert/M$, we obtain
\begin{equation}
\tilde{V}=\frac{V}{\kappa^2M^4}=(y^2-1-\xi y^4)^2+
2w^2y^2(1-2\xi y^2)^2, 
\label{eq:vtilde}
\end{equation}
where $\xi=\beta M^2/\kappa M_S^2$. This potential is a 
simple extension of the standard potential for SUSY hybrid 
inflation (which corresponds to $\xi=0$) and appears in a 
wide class of models incorporating the leading 
non-renormalizable correction to the standard hybrid 
inflationary superpotential.
 
For constant $w$ (or $|S|$), $\tilde V$ in 
Eq.(\ref{eq:vtilde}) has extrema at  
\begin{equation}
y_1=0,~y_2=\frac{1}{\sqrt{2\xi}},~y_{3\pm}=\frac{1}
{\sqrt{2\xi}}\sqrt{(1-6\xi w^2)\pm\sqrt{(1-6\xi w^2)^2
-4\xi(1-w^2)}}. 
\label{eq:extrema}
\end{equation}
Note that the first two extrema (at $y_1$, $y_2$) are 
$|S|$-independent and, thus, correspond to classically 
flat directions, the trivial one at $y_1=0$ with 
$\tilde{V}_1=1$, and the non-trivial one at  
$y_2=1/\sqrt{2\xi}={\rm constant}$ with 
$\tilde{V}_2=(1/4\xi-1)^2$, which we will use as our 
inflationary path. The trivial trajectory is a valley of 
minima for $w>1$, while the non-trivial one for 
$w>w_0=(1/8\xi-1/2)^{1/2}$, which is its instability 
(critical) point. We take $\xi<1/4$, so that $w_0>0$ and 
the non-trivial path is destabilized before $w$ 
reaches zero (the destabilization is in the chosen direction 
$H^c \,^{*}=\bar{H}^{c}$). The extrema at $y_{3\pm}$, 
which are $|S|$-dependent and non-flat, do not exist for 
all values of $w$ and $\xi$, since the expressions under 
the square roots in Eq.(\ref{eq:extrema}) are not always 
non-negative. These two extrema, at $w=0$, become the SUSY 
vacua. The relevant SUSY vacuum (see below) corresponds to 
$y_{3-}(w=0)$ and, thus, the absolute value $v_0$ of the 
common vev of $H^c \,^{*}$, $\bar{H}^{c}$ is given by
\begin{equation}
(\frac{v_0}{M})^2=\frac{1}{2\xi}(1-\sqrt{1-4\xi}).
\label{eq:v0}
\end{equation}

We will now discuss the structure of $\tilde{V}$ and the
inflationary history in the most interesting range of $\xi$, 
which is $1/4>\xi>1/6$. For fixed $w>1$,
there exist two local minima at $y_1=0$ and 
$y_2=1/\sqrt{2\xi}$, which corresponds to lower potential
energy density, and a local maximum at $y_{3+}$ lying 
between the minima. As $w$ becomes smaller than unity, the 
extremum at $y_1$ turns into a local maximum, while the 
extremum at $y_{3+}$ disappears. The system can freely fall 
into the non-trivial (desirable) trajectory at $y_2$ even 
if it started at $y_1=0$. As we further decrease $w$ below 
$(2-\sqrt{36\xi-5})^{1/2}/3\sqrt{2\xi}$, a pair of new 
extrema, a local minimum at $y_{3-}$ and a local maximum at 
$y_{3+}$, are created between $y_1$ and $y_2$. As $w$
crosses $(1/8\xi-1/2)^{1/2}$, the local maximum at 
$y_{3+}$ crosses $y_2$ becoming a local minimum. At the 
same time, the local minimum at $y_2$ turns into a local 
maximum and inflation along the `shifted' trajectory is 
terminated with the system falling into the local minimum at 
$y_{3-}$ which, at $w=0$, develops into a SUSY vacuum.

We see that, no matter where the system starts from, it 
always passes from the `shifted' trajectory, where the 
relevant part of inflation takes place, before falling 
into the SUSY vacuum. So, $G_{PS}$ is already broken 
during inflation and no monopoles are produced at the 
`waterfall'.

The COBE \cite{cobe} result can be reproduced, for instance, 
with $\kappa\approx 4\times 10^{-3}$, which corresponds to 
$\xi=1/5$, $v_0\approx 1.7\times 10^{16}~{\rm GeV}$, 
$M\approx 1.45\times 10^{16}~{\rm GeV}$ (for $\beta=1$, 
$M_S=5\times 10^{17}~{\rm GeV}$). Notice that 
$v_0\sim 10^{16}~{\rm GeV}$ consistently with the 
unification of the MSSM gauge couplings. The spectral index 
$n=0.954$.

After inflation, the system could possibly fall into the 
minimum at $y_{3+}$. This, however, does not happen since in 
the last e-folding or so the barrier between the minima at 
$y_{3-}$ and $y_2$ is considerably reduced and the decay of 
the `false vacuum' at $y_2$ to the minimum at $y_{3-}$ is 
completed within a fraction of an e-folding before the 
$y_{3+}$ minimum even comes into existence. 

\subsection{Smooth Hybrid Inflation}
\label{subsec:smooth}

An alternative solution \cite{smooth,jean2} to the monopole 
problem of hybrid inflation can be constructed by imposing, in 
the model of Sec.\ref{subsec:shifted}, an extra $Z_2$ symmetry 
under which $H^c\bar{H}^c\rightarrow -H^c\bar{H}^c$ (say 
$H^c\rightarrow -H^c$). The whole structure of the model 
remains unaltered except that now only even powers of the 
combination $H^c\bar{H}^c$ are allowed in the superpotential 
terms. 

The inflationary superpotential in Eq.(\ref{eq:susyinfl}) 
becomes
\begin{equation}
W=S\left(-\mu^2+\frac{(H^c\bar{H}^c)^2}
{M_S^2}\right), 
\label{eq:smoothsuper}
\end{equation}
where we absorbed the dimensionless parameters $\kappa$, 
$\beta$ in $\mu$, $M_S$. The resulting scalar potential
$V$ is then given by
\begin{equation}
\tilde{V}=\frac{V}{\mu^4}=(1-\tilde\chi^4)^2+
16\tilde\sigma^2\tilde\chi^6, 
\label{eq:smoothpot}
\end{equation}
where we used the dimensionless fields $\tilde\chi=
\chi/2(\mu M_S)^{1/2}$, $\tilde\sigma=
\sigma/2(\mu M_S)^{1/2}$ with $\chi$, $\sigma$ being 
normalized real scalar fields defined by 
$\nu_H^c=\bar{\nu}_H^c=\chi/2$, $S=\sigma/\sqrt{2}$ 
after rotating $\nu_H^c$, $\bar{\nu}_H^c$, $S$ to the 
real axis.

The emerging picture is completely different. The flat 
direction at $\tilde\chi=0$ is now a local maximum with
respect to $\tilde\chi$ for all values of $\tilde\sigma$,
and two new symmetric valleys of minima appear 
\cite{smooth,jean2} at
\begin{equation}
\tilde\chi=\pm\sqrt{6}\tilde\sigma\left[\left(1+
\frac{1}{36\tilde\sigma^4}\right)^{\frac{1}{2}}-1
\right]^{\frac{1}{2}}.
\label{eq:smoothvalley}
\end{equation}
They contain the SUSY vacua which lie at $\tilde\chi=
\pm 1$, $\tilde\sigma=0$. Note that these valleys are not 
classically flat. In fact, they possess an inclination already 
at the classical level, which can drive the inflaton towards 
the vacua. As a consequence, contrary to the case of standard 
SUSY or shifted hybrid inflation, there is no need of radiative 
corrections, which are expected to give a subdominant 
contribution to the slope of the
inflationary paths. In spite of this, one could try to include
the one-loop corrections. This requires the construction of the
mass spectrum on the inflationary trajectories. In doing so, we
find that the ${\rm mass}^2$ of some scalars belonging to the
inflaton sector is negative. The one-loop corrections, which
involve logarithms of the masses squared, are then ill-defined.
This may be remedied by resumming the perturbative expansion to
all orders, which is a formidable task and we do not pursue it  
here. 

The potential along the symmetric valleys of minima is given by 
\cite{smooth,jean2}
\begin{eqnarray}
\tilde{V}&=&48\tilde\sigma^4\left[72\tilde\sigma^4\left(1+
\frac{1}{36\tilde\sigma^4}\right)\left(\left(1+
\frac{1}{36\tilde\sigma^4}\right)^{\frac{1}{2}}-1\right)
-1\right]
\nonumber \\
&=&1-\frac{1}{216\tilde\sigma^4}+\cdots,~~{\rm for}
~\tilde\sigma\gg 1.
\label{eq:smoothV}
\end{eqnarray}
The system follows, from the beginning, a particular 
inflationary trajectory and, thus, ends up at a particular 
point of the vacuum manifold leading to no production of 
disastrous magnetic monopoles.

\par
Inflation does not come to an abrupt end in this case since 
the inflationary path is stable with respect to $\tilde\chi$ 
for all $\tilde\sigma$'s. The value $\tilde\sigma_0$ of 
$\tilde\sigma$ at which inflation is terminated smoothly is 
found from the $\epsilon$ and $\eta$ criteria (see e.g., 
Ref.\cite{cosmology}), and the derivatives \cite{jean2} of 
the potential along the inflationary path:

\begin{equation}
\frac{d\tilde{V}}{d\tilde\sigma}=192\tilde\sigma^3
\left[(1+144\tilde\sigma^4)\left(\left(1+
\frac{1}{36\tilde\sigma^4}\right)^{\frac{1}{2}}
-1\right)-2\right],
\label{eq:firstder}
\end{equation}
\begin{eqnarray}
\frac{d^2\tilde{V}}{d\tilde\sigma^2}&=&
\frac{16}{3\tilde\sigma^2}
\Biggl\{(1+504\tilde\sigma^4)
\left[72\tilde\sigma^4\left(\left(1+
\frac{1}{36\tilde\sigma^4}\right)^{\frac{1}{2}}
-1\right)-1\right]
\nonumber \\
& &-(1+252\tilde\sigma^4)\left(\left(1+
\frac{1}{36\tilde\sigma^4}\right)^{-\frac{1}{2}}
-1\right)\Biggl\}.
\label{eq:secondder}
\end{eqnarray}

Here, we have the freedom to identify the vev $v_0=
\vert\langle H^c\rangle\vert=\vert\langle\bar{H}^c\rangle
\vert$, which equals $(\mu M_S)^{1/2}$, with 
the SUSY GUT scale $M_G\approx 2.86\times 10^{16}~{\rm GeV}$. 
From COBE \cite{cobe}, we then obtain 
$M_S\approx 4.39\times 10^{17}~{\rm GeV}$ and 
$\mu\approx 1.86\times 10^{15}~{\rm GeV}$.

\section{Relevant Phenomenological and Cosmological 
Constraints: Shifted versus Smooth Hybrid Inflation}
\label{sec:constraints}

\subsection{The $\mu$ Problem}
\label{subsec:mu}

An important shortcoming of MSSM is that there is no 
understanding of how the SUSY $\mu$-term, with the right 
magnitude of $|\mu|\sim 10^{2}-10^{3}~{\rm GeV}$, arises. 
In both scenarios of shifted and smooth hybrid inflation, one 
way \cite{PQmu} to solve this $\mu$ problem is via a PQ 
symmetry $U(1)_{PQ}$ \cite{pq}, which also solves the strong 
CP problem. This solution is based on the observation \cite{kn} 
that the axion decay constant $f_{a}$, 
which is the symmetry breaking scale of $U(1)_{PQ}$, is 
(normally) `intermediate' ($\sim 10^{11}-10^{12}~{\rm GeV}$) 
and, thus, $|\mu|\sim f_{a}^2/M_S$. The scale $f_{a}$ is, 
in turn, $\sim (m_{3/2}M_S)^{1/2}$, where 
$m_{3/2}\sim 1~{\rm{TeV}}$ is 
the gravity-mediated soft SUSY breaking scale (gravitino mass). 
In order to implement this solution of the $\mu$ problem, we 
introduce a pair of gauge singlet superfields $N$, $\bar{N}$ 
with PQ charges -1, 1 and the non-renormalizable couplings 
$\lambda_1N^2h^2/M_S$, $\lambda_2N^2\bar{N}^2/M_S$ in 
the superpotential. Here, $\lambda_{1,2}$ are taken 
positive by redefining the phases of $N$, $\bar{N}$. After 
SUSY breaking, the $N^2\bar N^2$ term leads to the scalar 
potential:
\begin{eqnarray}
V_{PQ}&=&\left(m_{3/2}^2
+4\lambda_2^2\left|\frac{N\bar{N}}{M_S}\right|^2\right)
\left[(|N|-|\bar{N}|)^2+2|N||\bar{N}|\right]
\nonumber \\
& &+2|A|m_{3/2}\lambda_2\frac{|N\bar{N}|^2}{M_S}
{\rm{cos}}(\epsilon+2\theta+2\bar{\theta}),
\label{eq:pqpot}
\end{eqnarray} 
where $A$ is the dimensionless coefficient of the soft SUSY 
breaking term corresponding to the superpotential term 
$N^2\bar{N}^2$ and $\epsilon$, $\theta$, $\bar{\theta}$  
are the phases of $A$, $N$, $\bar{N}$ respectively. 
Minimization of $V_{PQ}$ then requires 
$|N|=|\bar{N}|$, $\epsilon+2\theta+2\bar{\theta}=\pi$ 
and $V_{PQ}$ takes the form
\begin{equation}
V_{PQ}=2|N|^2m_{3/2}^2\left(4\lambda_2^2\frac{|N|^4}
{m_{3/2}^2M_S^2}-|A|\lambda_2\frac{|N|^2}{m_{3/2}M_S}
+1\right).
\label{eq:pqpotmin}
\end{equation}
For $|A|>4$, the absolute minimum of the potential is at
\begin{equation}
|\langle N\rangle|=|\langle\bar{N}\rangle|\equiv
\frac{f_a}{2}=(m_{3/2}M_S)^{\frac{1}{2}}
\left(\frac{|A|+(|A|^2-12)^{\frac{1}{2}}}
{12\lambda_2}\right)^{\frac{1}{2}}\sim (m_{3/2}M_{S})
^{\frac{1}{2}}.
\label{eq:solution}
\end{equation}
The $\mu$-term is generated via the $N^2h^2$ superpotential 
term with $|\mu|=2\lambda_1|\langle N\rangle|^2/M_S$, 
which is of the right magnitude.

The potential $V_{PQ}$ also has a local minimum at 
$N=\bar{N}=0$, which is separated from the global PQ minimum by 
a sizable potential barrier preventing a successful transition 
from the trivial to the PQ vacuum. This situation persists at all 
cosmic temperatures after the `reheating' which follows hybrid 
inflation, as has been shown \cite{jean1} by considering the 
one-loop temperature corrections \cite{jackiw} to the potential. 
We are, thus, obliged to assume that, after the end of inflation, 
the system emerges in the PQ vacuum since, otherwise, it will be 
stuck for ever in the trivial vacuum.

\subsection{`Reheating' and Leptogenesis}
\label{subsec:reheat}

A complete inflationary scenario should be followed by a 
successful `reheating' satisfying the gravitino constraint 
\cite{khlopov} on the `reheat' temperature, $T_r
\stackrel{_{<}}{_{\sim}}10^9~{\rm GeV}$, and 
generating the observed BAU. After the end of inflation, 
the system falls towards 
the SUSY vacuum and performs damped oscillations about it. 
The inflaton (oscillating system) consists of the two 
complex scalar fields $\theta=(\delta\nu^c_H+
\delta\bar\nu^c_H)/\sqrt{2}$ ($\delta\nu^c_H=
\nu^c_H-v_0$, $\delta\bar\nu^c_H=\bar\nu^c_H-v_0$) and 
$S$, with equal mass $m_{\rm infl}=\sqrt{2}\kappa v_0
(1-2\xi v_0^2/M^2)$ or $2\sqrt{2}(\mu/M_S)^{1/2}\mu$ 
for shifted or smooth hybrid inflation respectively. 

The fields $\theta$ and $S$ decay into a pair of right-handed 
neutrinos ($\psi_{\nu^c_i}$) and sneutrinos ($\nu^c_i$) 
respectively via the coupling 
$\gamma_i\bar{H}^c\bar{H}^c F_i^cF_i^c/M_S$ 
and the terms in Eq.(\ref{eq:susyinfl}) or 
(\ref{eq:smoothsuper}) in the shifted or smooth case. The 
Lagrangian terms are:
\begin{equation}
L^\theta_{\rm decay}=-\sqrt{2}\gamma_i\frac{v_0}{M_S}
\theta\psi_{\nu_i^c}\psi_{\nu_i^c}+ h.c.~, 
\label{eq:thetadecay}
\end{equation}
\begin{equation}
L^S_{\rm decay}=-\sqrt{2}\gamma_i\frac{v_0}{M_S}S^*\nu^c_i
\nu^c_im_{\rm infl}+h.c.~,  
\label{eq:sdecay}
\end{equation}
and the common, as it turns out, decay width is given by
\begin{equation}
\Gamma=\Gamma_{\theta\rightarrow\bar\psi_{\nu^c_i}
\bar\psi_{\nu^c_i}}=\Gamma_{S\rightarrow\nu^c_i\nu^c_i}=
\frac{1}{8\pi}\left(\frac{M_i}{v_0}\right)^2m_{\rm infl}~,
\label{eq:gamma}
\end{equation}
provided that the mass $M_i=2\gamma_iv_0^2/M_S$ of the relevant
$\nu^c_i$ satisfies the inequality $M_i<m_{\rm infl}/2$. The 
same number of particles and sparticles is produced after 
inflation, and thus the SUSY world is recovered.

To minimize the number of small coupling constants, we assume 
that
\begin{equation}
M_2<\frac{1}{2}m_{\rm infl}\leq M_3~,
\label{eq:ineq}
\end{equation}
so that the coupling $\gamma_3$ can be of order unity. The 
inflaton then decays into the second heaviest right-handed 
neutrino superfield with mass $M_2$. Note that there always 
exist $\gamma_3$'s smaller than unity such that the second 
inequality in Eq.(\ref{eq:ineq}) is satisfied for all 
relevant values of the other parameters. 

The `reheat' temperature $T_r$, for the MSSM spectrum, is 
given \cite{lss} by
\begin{equation}
T_r\approx\frac{1}{7}(\Gamma M_P)^{\frac{1}{2}},
\label{eq:reheat}
\end{equation}
and must satisfy the gravitino constraint \cite{khlopov}, 
$T_r\stackrel{_{<}}{_{\sim}}10^9~{\rm GeV}$, for 
gravity-mediated SUSY breaking with universal boundary 
conditions. To maximize the naturalness of the model, we take 
the maximal $M_2$ (and thus $\gamma_2$) allowed by the 
gravitino constraint. These $M_2$'s turn out to be much 
smaller than the values of $m_{\rm{infl}}/2$ and, thus, the 
first inequality in Eq.(\ref{eq:ineq}) is well satisfied.

Another important constraint comes from the BAU. In this model, 
a primordial lepton asymmetry \cite{leptogenesis} is produced 
which is then partly converted into baryon asymmetry by the 
non-perturbative electroweak sphaleron effects \cite{sphaleron}. 
Actually, in the PS model under consideration as well as in 
many other models, this is the only way to generate the observed 
BAU since the inflaton decays into right-handed neutrino 
superfields. The subsequent decay of these superfields into 
lepton (antilepton) $L$ ($\bar L$) and electroweak Higgs 
superfields can only produce a lepton asymmetry. It is important 
to ensure that this lepton asymmetry is not erased \cite{turner} 
by lepton number violating $2 \rightarrow 2$ scattering 
processes such as $LL\rightarrow h_2^{*}h_2^{*}$ or 
$Lh_2\rightarrow\bar{L}h_2^{*}$ at all 
temperatures between $T_{r}$ and $100~{\rm GeV}$. This is 
automatically satisfied since the lepton asymmetry is 
protected \cite{ibanez} by SUSY at temperatures between 
$T_r$ and $T \sim 10^{7}~{\rm GeV}$ and, for 
$T\stackrel{_{<}}{_{\sim }}10^{7}~{\rm GeV}$, these 
scattering processes are well out of equilibrium provided 
\cite{ibanez} 
$m_{\nu_{\tau}}\stackrel{_<}{_\sim} 10~{\rm{eV}}$, 
which readily holds in our case (see below). For MSSM 
spectrum, the observed BAU $n_B/s$ is related \cite{ibanez} 
to the primordial lepton asymmetry $n_L/s$ by 
$n_B/s=(-28/79)n_L/s$. Thus, the low deuterium abundance 
constraint \cite{deuterium} on the BAU gives $1.8\times 
10^{-10}\stackrel{_<}{_\sim}-n_L/s\stackrel{_<}{_\sim} 
2.3\times 10^{-10}$. 

As already mentioned, the lepton asymmetry is produced 
through the decay of the superfield $\nu^{c}_{2}$, which 
emerges as decay product of the inflaton. This superfield 
decays into electroweak Higgs and (anti)lepton superfields. 
The relevant one-loop diagrams are both of the vertex and 
self-energy type \cite{covi} with an exchange of 
$\nu^{c}_{3}$. The resulting lepton asymmetry is 
\cite{Laz3}
\begin{equation}
\frac{n_{L}}{s}\approx 1.33~\frac{9T_{r}}
{16\pi m_{\rm infl}}~\frac{M_2}{M_3}
~\frac{{\rm c}^{2}{\rm s}^{2}\sin 2\delta
(m_{3}^{D}\,^{2}-m_{2}^{D}\,^{2})^{2}}
{|\langle h_2\rangle|^{2}~(m_{3}^{D}\,^{2} 
{\rm \ s}^{2}+m_{2}^{D}\,^{2}{\rm \ c^{2}})}~,
\label{eq:leptonasym}
\end{equation}
where $|\langle h_2\rangle|\approx 174~\rm{GeV}$, 
$m_{2,3}^{D}$ ($m_{2}^{D}\leq m_{3}^{D}$) are the 
`Dirac' neutrino masses (in a basis where they are diagonal 
and positive), and ${\rm c}=\cos\theta$,  
${\rm s}=\sin\theta$, with $\theta$ and $\delta$ being 
the rotation angle and phase which diagonalize the Majorana 
mass matrix of the right-handed neutrinos. Note that 
Eq.(\ref{eq:leptonasym}) holds \cite{pilaftsis} provided 
that $M_{2}\ll M_{3}$ and the decay width of 
$\nu^{c}_{3}$ is $\ll(M_{3}^{2}-M_{2}^{2})/M_{2}$, 
and both conditions are well satisfied in our model. Here, 
we concentrated on the 
two heaviest families ($i=2,3$) and ignored the first one. 
We were able to do this since the analysis \cite{giunti} of 
the CHOOZ experiment \cite{chooz} shows that the solar and 
atmospheric neutrino oscillations decouple. 

The light neutrino mass matrix is given by the seesaw 
formula:
\begin{equation}
m_{\nu}\approx-\tilde m^{D}\frac{1}{M}m^{D},
\label{eq:neumass}
\end{equation}
where $m^{D}$ is the `Dirac' neutrino mass matrix and 
$M$ the Majorana mass matrix of right-handed neutrinos.
The determinant and the trace invariance of 
$m_{\nu}^\dagger m_{\nu}$ imply \cite{Laz3} two 
constraints on the asymptotic (at $M_G$) parameters which 
take the form: 
\begin{equation}
m_{2}m_{3}=\frac{\left(m_{2}^{D}m_{3}^{D}\right)^{2}}
{M_{2}M_{3}}~,
\label{eq:determinant}
\end{equation}
\begin{eqnarray*}
m_{2}\,^{2}+m_{3}\,^{2}=\frac{\left(m_{2}^{D}\,^{2}
{\rm \ c}^{2}+m_{3}^{D}\,^{2}{\rm \ s}^{2}\right)^{2}}
{M_{2}\,^{2}}+
\end{eqnarray*}
\begin{equation}
\frac{\left(m_{3}^{D}\,^{2}{\rm \ c}^{2}+
m_{2}^{D}\,^{2}{\rm \ s}^{2}\right)^{2}}{M_{3}\,^{2}}+
\frac{2(m_{3}^{D}\,^{2}-m_{2}^{D}\,^{2})^{2}
{\rm c}^{2}{\rm s}^{2}{\cos 2\delta }}{M_{2}M_{3}}~,
\label{eq:trace} 
\end{equation}
where $m_{2}=m_{\nu_\mu}$ and $m_{3}=m_{\nu_\tau}$ are the
(positive) eigenvalues of $m_{\nu}$, which are restricted by 
the small or large mixing angle MSW solution \cite{bahcall} of 
the solar neutrino puzzle and SuperKamiokande \cite{japan} 
respectively.

The $\mu-\tau$ mixing angle $\theta_{23}=\theta _{\mu\tau}$ 
lies \cite{Laz3} in the range
\begin{equation}
|\,\varphi -\theta ^{D}|\leq \theta _{\mu\tau}\leq
\varphi +\theta ^{D},\ {\rm {for}\ \varphi +
\theta }^{D}\leq \ \pi /2~,
\label{eq:mixing}
\end{equation}
where $\varphi$ is the rotation angle which diagonalizes 
$m_{\nu}$ in the basis where $m^D$ is diagonal and 
$\theta^{D}$ is the `Dirac' mixing angle (i.e., the 
`unphysical' mixing angle with zero Majorana 
masses for the right-handed neutrinos). We will assume, for 
simplicity, that $\theta^D$ is negligible, which implies
$\theta_{\mu\tau}\simeq\varphi$. Also due to the presence 
of $SU(4)_c$ in $G_{PS}$, $m^D_3$ 
coincides with the asymptotic value of the top quark mass. 
Taking renormalization effects into account, in the context of 
the MSSM with large $\tan\beta$, we find \cite{Laz3} 
$m^D_3=110-120~{\rm{GeV}}$. We also include the running 
of $\theta_{\mu\tau}$ from $M_G$ to the electroweak scale 
\cite{springer}. 

In shifted hybrid inflation, for each $\kappa$ and $\gamma_3$, 
the $M_{2,3}$ are fixed. Taking $m_{2,3}$ and 
$m^D_3$ also fixed in their allowed ranges, we are left with 
three undetermined parameters $\delta$, $\theta$ and 
$m^D_2$ which are restricted by four constraints: 
almost maximal $\nu_{\mu}-\nu_{\tau}$ mixing 
($\sin^{2}2\theta_{\mu\tau}\stackrel{_>}{_\sim}0.85$) 
from SuperKamiokande \cite{japan}, the leptogenesis 
bound ($1.8\times 10^{-10} \stackrel{_<}{_\sim}
-n_L/s\stackrel{_<}{_\sim}2.3\times 10^{-10}$), and 
Eqs.(\ref{eq:determinant}) and (\ref{eq:trace}). It is 
highly non-trivial that solutions satisfying all the above 
requirements can be found with natural $\kappa$'s 
($\sim 10^{-3}$) and $m^D_2$'s of order $1~{\rm GeV}$ (see 
last paragraph of this section). Typical solutions can be 
constructed, for instance, for $\kappa=4\times 10^{-3}$ (see 
Sec.\ref{subsec:shifted}), which gives $m_{\rm{infl}}\simeq 
4.1\times 10^{13}~{\rm GeV}$, $M_2\simeq 5.9\times 10^{10}
~{\rm GeV}$ and $M_3\simeq 1.1\times 10^{15}~{\rm GeV}$ (for 
$\gamma_3=0.5$). Taking, for example, $m_{\nu_{\mu}}=7.6
\times 10^{-3}~{\rm eV}$, $m_{\nu_{\tau}}=8\times 10^{-2}
~{\rm eV}$ and $m^D_3=120~{\rm GeV}$, we find $m^D_2\simeq 
1.2~{\rm GeV}$, $\sin^2 2\theta_{\mu\tau}\simeq 0.9$,
$n_L/s\simeq -1.8\times 10^{-10}$ and $\theta\simeq 0.016$ 
for $\delta\simeq -\pi/3$. Note that the $m_{\nu_\mu}$'s, 
for which solutions are found, turn out to be consistent with 
the large rather than the small mixing angle MSW mechanism.

In smooth hybrid inflation, we observe that no solutions can be 
found with $T_r\stackrel{_{<}}{_{\sim }}10^9~{\rm GeV}$, 
which is the gravitino constraint as usually quoted. We thus take 
$T_r=10^{10}~{\rm GeV}$, which is also perfectly acceptable 
provided \cite{kawasaki} that the branching ratio of the 
gravitino to photons is somewhat smaller than unity and the 
gravitino mass is relatively large ($\sim$ a few hundred GeV).

\begin{figure}[ht]
\begin{center}
\hspace{-2.5cm}\psfig{figure=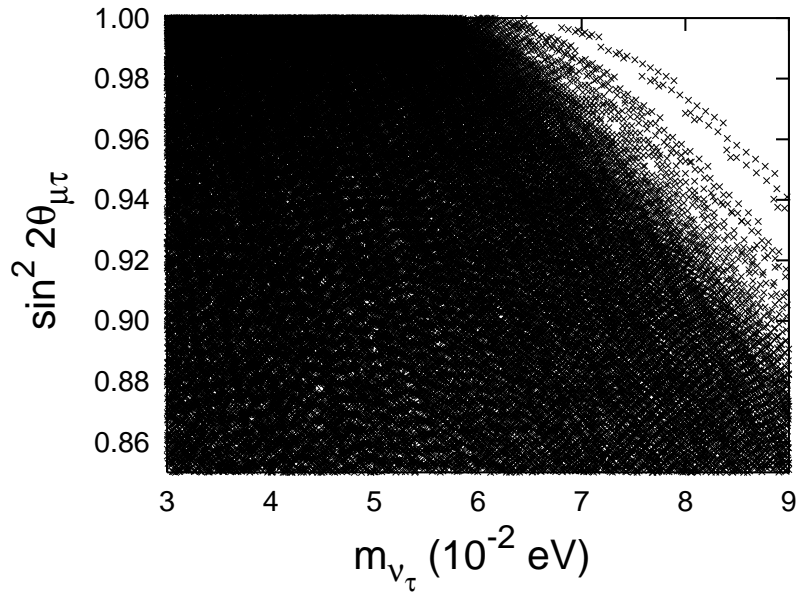,height=8cm}
\end{center}
\caption{The scatter plot in the $m_{\nu_\tau}-
\sin^2 2 \theta_{\mu\tau}$ plane of the solutions which 
satisfy the low deuterium abundance constraint on the BAU, the 
restrictions from solar and atmospheric neutrino oscillations, 
and the $SU(4)_c$ invariance in the case of smooth hybrid 
inflation. We take $T_r=10^{10}~{\rm GeV}$, $\gamma_3\approx 
0.05-0.5$, $m^D_2\approx 0.8-2~{\rm GeV}$ and $m^D_3\approx 
110-120~{\rm GeV}$.}
\label{sin2n1}
\end{figure}

Our results are shown in Fig.\ref{sin2n1}, where we plot 
solutions corresponding to $T_r=10^{10}~{\rm GeV}$ and 
satisfying the leptogenesis constraint consistently with the 
neutrino oscillation data and the $SU(4)_c$ symmetry. The 
parameter $\gamma_3$ runs from $0.05$ to $0.5$, i.e., 
$M_3\approx 1.86\times 10^{14}-1.86\times 10^{15}
~{\rm GeV}$. The second inequality in Eq.(\ref{eq:ineq}) 
implies that $\gamma_3\stackrel{_{>}}{_{\sim }}0.046$. 
However, no solutions are found for $\gamma_3<0.05$. Also, 
values of $\gamma_3$ higher than 0.5 do not allow solutions. 
The mass of the second heaviest right-handed neutrino 
$M_2\approx 1.55\times 10^{11}~{\rm GeV}$, which clearly 
satisfies the first inequality in Eq.(\ref{eq:ineq}). The 
restrictions from $SU(4)_c$ invariance are expected to be 
more or less accurate only if applied to the masses of the 
third family quarks and leptons. For the second family, they 
should hold only as order of magnitude relations. We thus 
restrict ourselves to values of $m^D_2$ smaller than 
$2~{\rm GeV}$ since much bigger $m^D_2$'s would violate 
strongly the $SU(4)_c$ symmetry (the value of $m^D_2$ from 
exact $SU(4)_c$ is \cite{Laz3} about $0.23~{\rm GeV}$ for 
MSSM spectrum with large $\tan\beta$). Moreover, we find 
that solutions exist only if 
$m^D_2\stackrel{_{>}}{_{\sim }}0.8$. So we take 
$m^D_2\approx 0.8-2~{\rm GeV}$ and, 
as required by $SU(4)_c$ invariance, $m^D_3\approx 110-120
~{\rm GeV}$. Also, the phase 
$\delta\approx (-\pi/8)-(-\pi/5)$ and the rotation angle 
$\theta\approx 0.01-0.03$ for solutions to appear. Note that 
$\delta$'s close to 0 or $-\pi/2$ are excluded since they 
yield very small primordial lepton asymmetry. 

\section{Conclusions}

We presented, in the context of the PS SUSY GUT model, two 
`natural' extensions of hybrid inflation, which solve the 
cosmological monopole problem. These models reproduce the 
COBE measurements with `natural' values of the parameters 
and a PS breaking scale close to (or equal with) the SUSY 
GUT scale. A PQ symmetry is used to generate the $\mu$-term 
of MSSM and proton is practically stable. Inflation is 
followed by a successful `reheating' satisfying the gravitino 
constraint on the `reheat' temperature and generating the 
observed BAU via a primordial leptogenesis consistently with 
the requirements from solar and atmospheric neutrino 
oscillations and the $SU(4)_c$ symmetry.

\section*{Acknowledgement}
This work was supported by European Union under the TMR contract
ERBFMRX-CT96-0090 and the RTN contracts HPRN-CT-2000-00148 and
HPRN-CT-2000-00152. One of us (S. K.) was supported by PPARC.

\def\anj#1#2#3{{\it Astron. J.}~{\bf #1}~(#2)~#3}
\def\apj#1#2#3{{\it Astrophys. Journal}~{\bf #1}~(#2)~#3}
\def\apjl#1#2#3{{\it Astrophys. J. Lett.}~{\bf #1}~(#2)~#3}
\def\baas#1#2#3{{\it Bull. Am. Astron. Soc.}~{\bf #1}~(#2)~#3}
\def\cmp#1#2#3{{\it Commun. Math. Phys.}~{\bf #1}~(#2)~#3}
\def\grg#1#2#3{{\it Gen. Rel. Grav.}~{\bf #1}~(#2)~#3}
\def\jetpl#1#2#3{{\it JETP Lett.}~{\bf #1}~(#2)~#3}
\def\jetpsp#1#2#3{{\it JETP (Sov. Phys.)}~{\bf #1}~(#2)~#3}
\def\jhep#1#2#3{{\it JHEP}~{\bf #1}~(#2)~#3}
\def\jpa#1#2#3{{\it J. Phys.}~{\bf A~#1}~(#2)~#3}
\def\mnras#1#2#3{{\it Mon. Not. Roy. Astr. Soc.}
~{\bf #1}~(#2)~#3}
\def\n#1#2#3{{\it Nature}~{\bf #1}~(#2)~#3}
\def\npb#1#2#3{{\it Nucl. Phys.}~{\bf B~#1}~(#2)~#3}
\def\pl#1#2#3{{\it Phys. Lett.}~{\bf #1~B}~(#2)~#3}
\def\plb#1#2#3{{\it Phys. Lett.}~{\bf B~#1}~(#2)~#3}
\def\pr#1#2#3{{\it Phys. Reports}~{\bf #1}~(#2)~#3}
\def\prd#1#2#3{{\it Phys. Rev.}~{\bf D~#1}~(#2)~#3}
\def\prl#1#2#3{{\it Phys. Rev. Lett.}~{\bf #1}~(#2)~#3}
\def\prsla#1#2#3{{\it Proc. Roy. Soc. London}
~{\bf A~#1}~(#2)~#3}
\def\ptp#1#2#3{{\it Prog. Theor. Phys.}~{\bf #1}~(#2)~#3}
\def\spss#1#2#3{{\it Sov. Phys. -Solid State}
~{\bf #1}~(#2)~#3}
\def\ibid#1#2#3{{\it ibid.}~{\bf #1}~(#2)~#3}
\def\stmp#1#2#3{{\it Springer Trac. Mod. Phys.}
~{\bf #1}~(#2)~#3}

\end{document}